\def\year{2021}\relax
\documentclass[letterpaper]{article}
\usepackage{aaai21}
\usepackage{times}
\usepackage{helvet}
\usepackage{courier}

\usepackage{natbib}
\usepackage{caption}

\usepackage[hyphens]{url}
\usepackage{graphicx}
\graphicspath{ {figures/} }

\usepackage{tabularx}
\usepackage{amsmath}
\usepackage{subcaption}
\usepackage{booktabs}

\title{Out of the Shadows: Analyzing Anonymous' Twitter Resurgence during the 2020 Black Lives Matter Protests}
\author{Keenan Jones, Jason R.~C.~Nurse, Shujun Li\\}
\affiliations{
Institute of Cyber Security for Society (iCSS) \& School of Computing\\ 
University of Kent, UK \\
\{ksj5, j.r.c.nurse, s.j.li\}@kent.ac.uk}

\begin{document} 

\maketitle 

\begin{abstract} 
Recently, there had been little notable activity from the once prominent hacktivist group, Anonymous. The group, responsible for activist-based cyber attacks on major businesses and governments, appeared to have fragmented after key members were arrested in 2013. In response to the major Black Lives Matter (BLM) protests that occurred after the killing of George Floyd, however, reports indicated that the group was back. To examine this apparent resurgence, we conduct a large-scale study of Anonymous affiliates on Twitter. To this end, we first use machine learning to identify a significant network of more than 33,000 Anonymous accounts. Through topic modelling of tweets collected from these accounts, we find evidence of sustained interest in topics related to BLM. We then use sentiment analysis on tweets focused on these topics, finding evidence of a united approach amongst the group, with positive tweets typically being used to express support towards BLM, and negative tweets typically being used to criticize police actions. Finally, we examine the presence of automation in the network, identifying indications of bot-like behavior across the majority of Anonymous accounts. These findings show that whilst the group has seen a resurgence during the protests, bot activity may be responsible for exaggerating the extent of this resurgence.
\end{abstract}

\section{Introduction}

The hacktivist collective Anonymous has been absent in recent years. As noted in various articles~\cite{Uitermark2017}, the group appeared to have fragmented around 2013, when a number of high profile affiliates were arrested. Since this time the group has been relatively quiet, with few actions attributed to them. Additionally, whilst a large number of Anonymous affiliate accounts exist on the social media site Twitter, previous work found that the majority of accounts were no longer active as of December, 2019~\cite{Jones2020}. 
After the killing of George Floyd by a Minneapolis police officer on 25th May 2020, protests –- largely organized by the Black Lives Matter (BLM) movement -- erupted throughout the US. Alongside these protests, media outlets began to issue reports indicating that Anonymous activity had `surged'~\cite{Independant2020}, conducting operations to support the protests~\cite{AJC2020}. In tandem with this, Anonymous has seen an apparent rise in popularity on Twitter, with approximately 3.5 million additional accounts following @YourAnonNews, one of the most prominent Anonymous accounts~\cite{Independant2020}. Alongside this, however, there have been criticisms regarding both the legitimacy of this resurgence~\cite{Independant2020} and the degree to which it represents genuine support, rather than efforts by the group to reclaim lost relevancy~\cite{Telegraph2020}.

Relative to these reports, our paper presents the first study of this apparent Anonymous Twitter resurgence; examining the degree to which a resurgence has genuinely occurred and examining its possible causes. To this, we aim to answer the following research questions:

\begin{itemize}
\item \textbf{RQ1}: Has the Anonymous Twitter network seen an subtantial increase in new members during the 2020 BLM protests?

\item \textbf{RQ2}: Do the topics discussed by Anonymous Twitter accounts during this period indicate an interest in BLM and wider issues of policing and racial injustice?

\item \textbf{RQ3}: Is interest in these topics sustained after the period of increased BLM protest?

\item \textbf{RQ4}: Does the tone in which BLM-related tweets are discussed change perceptibly after significant events in the protest timeline, and what can we learn from these tweets about Anonymous' position in regards to the BLM protests?

\item \textbf{RQ5}: To what degree can any increase in activity within the Anonymous Twitter network be attributed to the use of automated accounts?
\end{itemize}

In answering these questions, our research finds considerable evidence that the group has received unprecedented growth in the months surrounding the 2020 BLM protests, and offers further computational analysis linking this resurgence to the BLM protests. To achieve this, we:

\begin{itemize}
\item Conducted time-series analysis to investigate growth in the Anonymous network during 2020 BLM protests. In turn, identifying considerable growth in the network over the protest period (RQ1).

\item Implemented topic modelling on Anonymous tweets posted during the protests. This led to the discovery of a number of topics related to police protest, issues of racial injustice, and other BLM-related subjects (RQ2).

\item Used our trained topic model to study interest in BLM tweets over time, revealing a large peak in interest shortly after George Floyd's death and little sustained interest after the period of sustained protest in June, 2020 (RQ3).

\item Carried out sentiment analysis to examine how the tone of tweets shifted over the protest period. This identified evidence of shifts in sentiment after George Floyd's death, with an increase in positive tweets in support of BLM. Further analysis of tweets before, during, and after the period of sustained BLM protest also identified a consistent unity in the group's support of BLM (RQ4).

\item Used bot-detection methods to identify the degree of automation present in Anonymous Twitter accounts. This exposed a high level of bot-like behavior across the Anonymous Twitter network, particularly for accounts tweeting about BLM (RQ5).
\end{itemize}

Our research provides quantitative evidence that the Anonymous group has resurged around the time of the BLM protests. Moreover, we find that this high level of interest in BLM appears to have waned very quickly, supporting criticisms in the media that this support was noncommittal~\cite{Telegraph2020}. Additionally, through our study of automation within the network we observed significant indication that this growth may have been artificial in nature via the use of bot accounts to inflate the group's presence.

\section{Background}
\label{Background}

\subsection{The Anonymous Collective}

Anonymous' first notable action began in 2007, when members of the controversial website 4Chan took action against the Church of Scientology, protesting the Church's apparent acts of online censorship~\cite{Olson2013}. To this, 4Chan members launched a series of actions against the Church, culminating in the use of cyber attacks against the Church's websites~\cite{Olson2013}.

After this, members of the group -- labelling themselves `Anonymous' -- began to launch further actions, typically centered around political activism~\cite{Olson2013}. These campaigns, or ``Ops'', included attacks against the Recording Industry Association of America for their actions against The Pirate Bay and against PayPal and Visa after their refusal to process Wikileaks donations~\cite{Uitermark2017}.

In 2011, a small number of prominent Anonymous members fractured to form the splinter group, LulzSec~\cite{Jones2020}. Under this new name, these members conducted a series of high-profile cyber attacks, including the leaking of account data from Sony Pictures, and the computer security firm Stratfor~\cite{Olson2013}.

The high-profile nature of LulzSec's actions soon culminated in the arrest of many of the group's affiliates. After these arrests, studies indicated that Anonymous, having lost its central figures, fragmented~\cite{Olson2013}. Since then, studies indicate that the group has been largely inactive~\cite{Jones2020,Uitermark2017}. 

\subsection{Black Lives Matter \& Anonymous} 

Black Lives Matter (BLM) is an activist movement founded in 2013 after the killing of black teenager Trayvon Martin and the following `not guilty' verdict of his killer~\cite{Carney2016}. After the killing of George Floyd in May 2020, the group saw a significant increase in support with large protests erupting throughout the USA~\cite{NYT2020}.

Coming out in support of these protests were a number of notable groups. This included an apparently resurged Anonymous collective~\cite{Independant2020}; a group with noted links to the BLM cause having engaged in actions both supporting~\cite{IBT2016} and attacking the movement in the recent past~\cite{Verge2016}. An inconsistency likely the result of the decentralized, swarm-like philosophies of the group~\cite{Uitermark2017}.

Perhaps owing to Anonymous' decentralized nature~\cite{Uitermark2017}, many of these attributed actions, including the group's apparent resurgence on Twitter~\cite{Independant2020}, have been called into question~\cite{Telegraph2020}. Particularly, the unlikely increase in Twitter activity, with some Anonymous accounts gaining millions of followers in the matter of months, has lead to suspicions regarding the veracity of the group's resurgence and, by extension, their support of the BLM movement~\cite{Telegraph2020}.

\section{Related Work}

In the past, there have been a few attempts to analyze the presence of Anonymous on Twitter. In~\cite{Beraldo2017}, the author uses social network analysis to study the proliferation of ``\#Anonymous'' on Twitter between 2012 and 2015. In particular, the author focuses on the stability of the network, examining the rate at which accounts tweeting  ``\#Anonymous'' remain in the network in subsequent months. In turn, the author found that stability was very low with the majority of accounts failing to appear in the network in consecutive months.

Moving beyond this, \citet{McGovern2020} narrowed the study of users tweeting ``\#Anonymous'', focusing specifically on the role of gender within these accounts. The authors proceeded to analyze the variety of `Ops' mentioned by both male and female members. The authors found that there was a clear distinction in the Ops being broadcast by Anonymous affiliates of different genders, with male affiliates focusing on a wide range of Ops including \#OpDeathEaters targeted at pursuing accused rapists and pedophiles and \#opsafewinter which sought to provide aid to the homeless during winter. In contrast, female Anonymous affiliates showed a more narrow set of interests typically focused on Ops dedicated towards preventing animal cruelty, such as \#OpKillingBay and \#opseaworld.

Moving beyond this research, \citet{Jones2020} shifted the focus away from accounts tweeting ``\#Anonymous'' to analyze the network structure of accounts that were explicitly affiliated with Anonymous. Using machine learning classification, the authors found that the group had a large presence on Twitter, with over 20,000 accounts being identified. Network centrality analysis was then used to identify patterns of influence in the group. From this, the authors found that, contrary to the group's claims of being leaderless and swarm-like, within the Twitter network there were a clear set of central influencer accounts. Moreover, the authors examined the evolution of the network over time, finding that the majority of the Anonymous accounts identified were no longer active as of 2019. A finding which supports statements in the literature that the group as a whole is largely inactive~\cite{Uitermark2017} .

\section{Our Contributions}

This paper builds upon prior research from the literature to offer the first study examining the resurgence of Anonymous on Twitter. In~\citep{Jones2020}, the authors conclude that Anonymous no longer appears to be active on Twitter. In this paper, we move beyond their work, which primarily focused on the structural properties of the group on Twitter, to present novel findings analyzing a significant resurgence in Anonymous activity, with more than 10,000 new Anonymous accounts joining the network in the first six months of 2020. This represents a significant increase on the 20,000 strong network found in \citep{Jones2020}.

Additionally, we utilize topic modelling and sentiment analysis to examine the tweeting habits of the Anonymous Twitter network at large, using these tools to draw a link between the Anonymous resurgence, and the rise in BLM protests after the death of George Floyd on 25th May, 2020. This is the first time, to our knowledge, that large-scale topic modelling and sentiment analysis of Anonymous tweets has been conducted. This reveals new insights into the topics of discussion present in tweets throughout the Anonymous network, with previous research only focusing on the tweeting habits of a small subset of 6 accounts~\cite{Jones2020}.

We also present the first use of topic analysis over time within this network, identifying that although a clear link between BLM and Anonymous' resurgence exists, this support appears to be short-lived. This new finding provides evidence to support media criticism that Anonymous BLM support is largely an attempt to reclaim relevance. 

Additionally, this paper conducts novel research into an aspect of Anonymous' Twitter presence that has received no prior investigation: the degree to which this presence is the result of bot activity. In turn, we show that the majority of new accounts joining the Anonymous network exhibit signs of automation, calling in to question the true degree to which the group's growth on Twitter constitutes a genuine resurgence. These results also raise doubts on the true extent of the Anonymous presence prior to 2020 reported in~\cite{Jones2020}, providing new insights indicating that the size of Anonymous' Twitter presence may have been driven by automated accounts.

\section{Methodology}
\label{Method}

In order to achieve this paper's aims of examining the apparent resurgence in the Anonymous Twitter network during the renewed BLM protests, we conducted a series of different computational methods to offer a detailed picture of the group's activity during that period and answer our RQs. We detail our process below:

\subsection{Data Collection}

In order to identify any apparent resurgences in the Anonymous Twitter network, it was first necessary to identify a large sample of Anonymous-affiliated accounts.

We opted to follow a similar approach to previous studies of the group~\cite{Jones2020} for our data gathering. By drawing on pre-existing sampling methods, we ensure that our data gathering process is robust. Moreover, we gain the ability to more accurately compare our sample of the (apparently resurged) Anonymous network at present to the findings made in past research.

We began with five notable Anonymous accounts, which would act as seeds from which the rest of the network could be sampled. These five accounts were drawn from a previous article that identified them as being notably associated with Anonymous~\cite{Jones2020}. Since this article was published, one of these accounts has been suspended, leaving four viable accounts. To this, we added an additional seed, `@YourAnonCentral', that had been recently identified in the media as a prominent Anonymous account~\cite{AJC2020}. 
We then conducted two-stage snowball sampling. For each seed account, all Anonymous-affiliated followers and friends (users that a given user follows)  were extracted using the Twitter Standard API~\cite{TwitterAPI}. This process was repeated for a second stage, for each of the followers and friends of the newly identified Anonymous accounts. 

In order to identify Anonymous accounts at each sampling stage, machine learning classification was used. To train the classifier, a subsection of the followers and friends of the five seed accounts were manually annotated as Anonymous or not Anonymous, using the strict definition of what constitutes an Anonymous account found in \cite{Jones2020}. In line with this definition, an account is annotated as `Anonymous' if it has at least one Anonymous-related keyword (Table~\ref{table:keywords}) in either its username or screen-name, and in its description, as well as having a profile or cover image containing either a Guy Fawkes mask or a floating businessman, images closely associated with the group~\cite{Olson2013}.

\begin{table}[!tbh]
\footnotesize
\centering
\begin{tabular}{ c c c c }
 \toprule
 anonymous & an0nym0u5 & anonymou5 & an0nymous \\
 anonym0us & anonym0u5 & an0nymou5 & an0nym0us \\ 
 anony & an0ny & anon & an0n\\ 
 legion & l3gion & legi0n & le3gi0n\\
     leg1on & l3g1on & leg10n & l3g10n\\
 \bottomrule
\end{tabular}
\caption{Anonymous keywords used.}
\label{table:keywords}
\end{table}

In turn, three annotators with expert knowledge of the group began by annotating a random sample of 200 accounts. Fleiss' Kappa agreement was then calculated between the three annotators, yielding a near-perfect agreement score of 0.92. Given the high level of agreement, a single annotator then proceeded to annotate the remaining accounts.

\begin{figure}[!ht]
\centering
\includegraphics[width=0.9\linewidth]{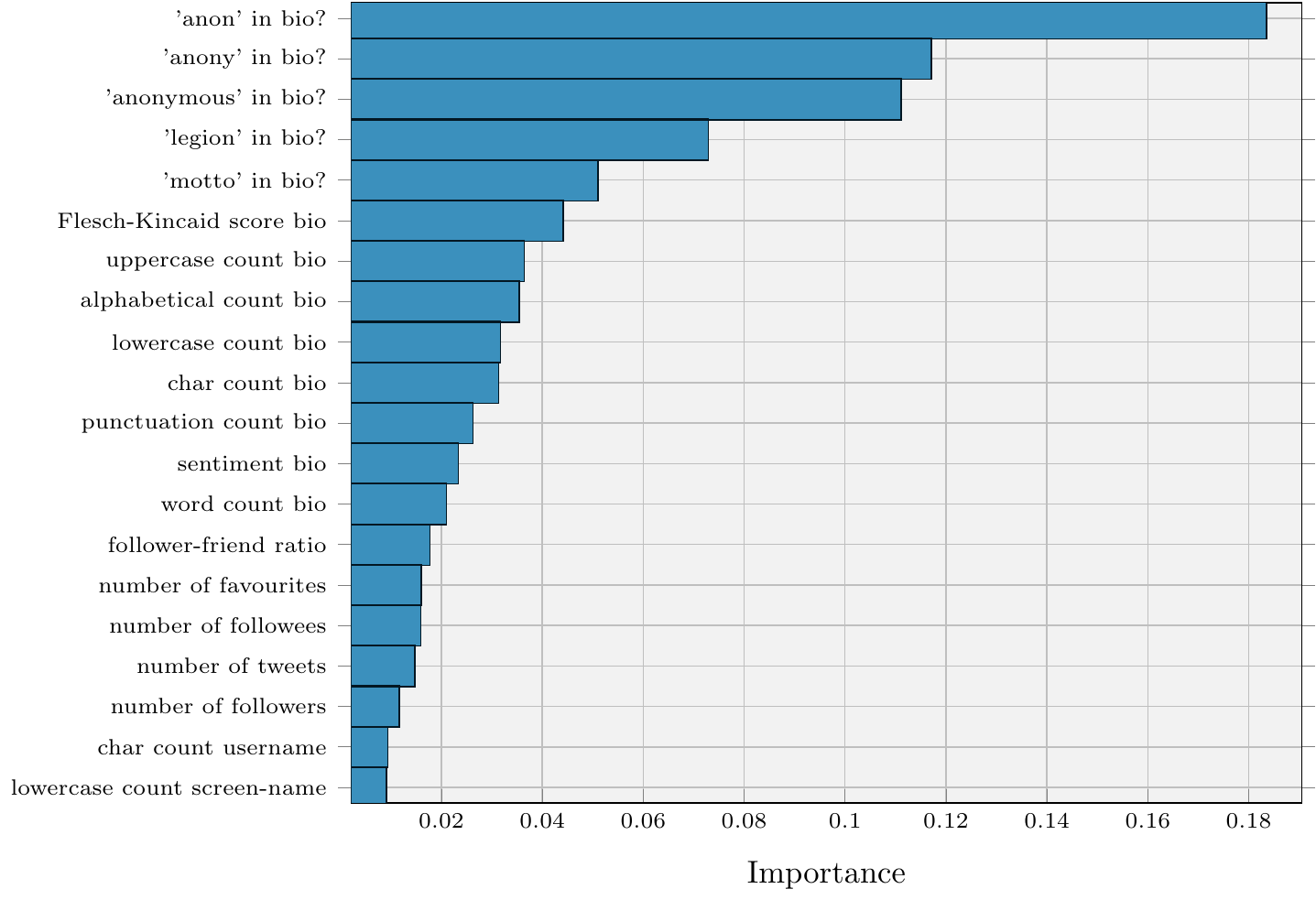}
\caption{Average feature importance for the top 20 most importance features for random forest across 100 trees.}
\label{fig:featureImportance}
\end{figure}

In total, 44,914 accounts were annotated, identifying 11,349 Anonymous accounts and 33,565 non-Anonymous accounts. This data was then used to train a series of classification algorithms -- random forest (RF), decision trees, and support vector machines (SVM) -- to identify the best performer. The results can be found in Table~\ref{table:ML_performances}, with all scores obtained using 5-fold cross validation and Scikit-learn's implementations of each algorithm~\cite{Pedregosa2011}. We opted to use the 62 account-based features adopted in \cite{Jones2020} given their proven efficacy in identifying Twitter accounts affiliated with specific groups. As RF was found to be the best performer this was chosen to identify Anonymous accounts at each sampling stage. The top 20 most important features for the RF classifier can be found in Fig.~\ref{fig:featureImportance}. Unsurprisingly, given the definition used at the annotation stage, the presence of Anonymous keywords and the Anonymous motto prove to be the most useful features. For clarity, all Anonymous keyword features were obtained using full token matching.

\begin{table}[!tbh]
\footnotesize
\centering
\begin{tabular}{ c c c c }
\toprule
\textbf{Model} & \textbf{Precision} & \textbf{Recall} & \textbf{F1-Score} \\
\midrule
Random forest & \textbf{0.94} & \textbf{0.94} & \textbf{0.94} \\ 
Decision tree & 0.91 & 0.91 & 0.91\\ 
SVM (sigmoid kernel) & 0.67 & 0.74 & 0.67\\
\bottomrule
\end{tabular}
\caption{Performances of the three tested machine learning models.}
\label{table:ML_performances}
\end{table}

In turn, our trained classifier was used to identify Anonymous accounts within the followers and friends of our five seed accounts (as of 31st September 2020). The model identified 31,562 accounts in this first stage. The followers and friends of these accounts were then collected, and the classifier used to identify Anonymous accounts in this new set of accounts. This process identified an additional 11,013 accounts, bringing our total network to 42,575.

As `anonymous', `anony', and `anon' are the top three most important features used by the classifier -- three terms with alternative meanings unrelated to the Anonymous collective -- there was the potential for misclassification that could affect further analysis. To ensure this had not occurred, we examined the Anonymous accounts identified by the classifier for the presence of false positives. Initially, we selected 10\% (4,258 accounts) of the classified accounts and identified the number of accounts within this set that did not clearly affiliate themselves with Anonymous. This identified that 80\% of these accounts had been identified correctly. Although this level of accuracy is not unreasonable, especially given the hard to define nature of the group, to ensure the reliability of our results we opted to extend this validation to the entire set of identified accounts. The total accuracy at the end of this validation remained steady at 79.19\%. We then removed all accounts that shared no affiliation with the group, leaving a final set of 33,820 Anonymous accounts. 

Within this set, there is still some argument to be made regarding the legitimacy of each account's affiliations with the group. However, given that Anonymous explicitly defines itself as having no set membership, with self-identification being sufficient to join the group~\cite{Olson2013}, we argue that a Twitter account's self identification is sufficient for it to be considered a `true' member.

Finally, as our research relied heavily on the analysis of tweet data posted by these Anonymous accounts, we used Twitter's Standard API~\cite{TwitterAPI} to extract the timelines from the newly identified network. The capabilities afforded to us by the API allowed for the extraction of the latest 3,200 tweets from each Anonymous account (as of 3rd December, 2020). 

As we are primarily interested in Anonymous' behavior over the period surrounding the 2020 BLM protests, we then removed all tweets posted before 1st May, 2020 and after 31st August, 2020. This date range was chosen via consultation with news articles discussing BLM actions which indicated that the protests began after George Floyd's death (on May 25th 2020)~\cite{AlJazeera2020, Independant2020} and peaked in activity in June, 2020~\cite{NYT2020}. This is also the period in which various actions attributed Anonymous occurred~\cite{AJC2020}. This date range allowed us to examine the permanence of any possible BLM association, and to see whether any association found existed before the protests began. Our timeline collection yielded approximately 7 million tweets in total. After filtering for date-range we were then left with 557,546 tweets.

\subsection{Identifying Anonymous' Topics of Interest}

In order to establish whether any surge in the Anonymous Twitter network could be attributed to the BLM protests (in answer to RQs 2 and 3), we conducted topic modelling on our corpus of Anonymous tweets. This is an approach that allowed us to identify the common topics of interest present across the Anonymous network during the protests.

Topic modelling refers to the use of unsupervised statistical models that attempt to learn the latent topics present in a collection of documents. LDA, the algorithm used in this paper, is among the most popular methods and operates under the assumption that a document is comprised of a series of latent topics~\cite{Kigerl2018}. The model uses probabilistic assignments of terms to a user specified number of topics. From this, each unique term in the corpus is assigned a probability distribution relative to the number of topics, indicating for each topic the probability that the term occurs within it. From this, the model can then be used to provide a distribution of topics over documents~\cite{Kigerl2018}.

As LDA requires the user to specify the number of topics, $k$, a value that was unknown in our research, it was necessary to utilize metrics to identify $k$'s optimum value. To this, we used \textit{CV} and UMass coherence~\cite{Syed2017}. These metrics assess the quality of $k$ by examining the semantic similarity between the top terms in each identified topic.

We then combined LDA topic modelling with the above metrics to identify the optimum set of topics within our Anonymous tweet corpus. To do this, we first processed our tweets: removing stopwords, short tweets ($<5$ tokens), and Twitter-specific noise (e.g., `RT' for retweet etc.), expanding contractions, and lemmatizing our data. Moreover, since the methods of data analysis we employed are only proven on texts written in English, any non-English tweets were removed from our dataset. This yielded a dataset of 189,781 tweets across 7,968 accounts.

Moving forward, the two coherence metrics were computed on topic models built from increasing values of $k$, starting at 5 and incrementing in steps of 5 to 50 topics. The optimum value for $k$ was then selected based on which value achieved the highest coherence scores. This was combined with a manual assessment that examined the degree to which optimum values of $k$ returned the most interpretable results.

As there have been indications that concatenating short documents can improve model performance~\cite{Kigerl2018}, we experimented with both single tweet documents and author documents (where each document consisted of all the tweets from a given account). We found there was little difference in the coherence scores achieved by the two approaches, nor any meaningful difference in the qualitative interpretability of the topics produced. Therefore, given the added flexibility of being able to compute the probability of topic occurrence at a more granular level using single tweet documents, this is the method we opted for. This approach identified 20 as the optimum topic number. For all topic modelling runs, we used LDA with a Gibbs sampler, using the Python Gensim wrapper for the MALLET LDA implementation~\cite{Gensim}.

We then trained our final LDA model, using the optimum topic number 20. The top terms identified for each topic were then examined manually and labelled according to the subject they likely represented. Our LDA model was then used to identify topics in each tweet in the corpus, using the distribution of topics over documents. A topic was considered present in a tweet if the probability of the topic occurring was greater than 0.8. 

\subsection{Analyzing Tweet Tone and Response}

Having identified the likely topics present in our dataset, we then wanted to identify the tone adopted in each tweet relative to key events in the 2020 BLM protest timeline. Particularly, before, during, and after the period between the death of George Floyd and the Juneteenth holiday (the last day of significant protest in the US~\cite{NYT2020}).

To do this, we opted to use VADER sentiment analysis (\url{https://github.com/cjhutto/vaderSentiment}), a popular tool optimized for use on social media posts~\cite{Gilbert2014}. We first applied VADER to each of the tweets used at the topic modelling stage. We then analyzed the average sentiment scores over the time-span captured within our tweets, for tweets belonging to specific topics of interest. This combined analysis allowed us to examine the response of tweets dedicated to given topics, as they related to (then) ongoing events in the BLM protests.

The sentiment scores in each time period were then compared to answer RQ4, using an manual analysis of tweets to provide additional context. Thereby providing an analysis of the tonal behaviors and changes in Anonymous tweets during each period and the degree to which these tonal behaviors indicated consensus, or lack thereof, within the Anonymous network.

\subsection{Detecting Bots in the Anonymous Network}

The final aim of our paper, in answer to RQ5, is to investigate the degree to which bot accounts were responsible for any substantial resurgence in the Anonymous Twitter network. 

To this end, we used the Botometer API (\url{https://botometer.iuni.iu.edu/}) to generate scores for each account in our Anonymous network, indicating the degree to which they exhibit automated behavior. Although there have been criticisms regarding the use of Botometer, we believe that it can be used in a legitimate manner, providing that the authors are careful to acknowledge the tool's limitations. 

In most criticisms of the tool~\cite{Rauchfleisch2020}, the research tests the quality of Botometer as means of classifying bot accounts, setting a score threshold designating an account as human, or bot. This is problematic as the authors do not recommend the use of Botometer in this way~\cite{Ferrara2020}. Instead, the tool advocates for the use of the score as the degree to which an account displays bot-like behavior. This is a more reasonable approach to take, examining the distribution of bot-like degree across the applied data as a whole to gain a sense of how bot-like a network is at scale.

We therefore used Botometer to provide bot scores for our identified Anonymous accounts. As we were limited to accounts tweeting in English, due to the tool's limitations in processing accounts tweeting in other languages, only accounts with English tweets were included. This resulted in bot scores being obtained for 27,059 Anonymous Twitter accounts (of the total 33,820).

\subsection{Ethics}

All data extractions were made in accordance with Twitter's Standard API terms and conditions~\cite{TwitterDev}, with no deleted, protected, or suspended accounts included in our analysis. Additionally, we do not name any accounts that have not already been named in existing articles. Finally, any tweet quotations included in this paper have been appropriately redacted/edited to ensure that their original author cannot be identified. These tweets have also been selected from Anonymous accounts that do not provide any personal/identifying information as an additional safeguard to user privacy.

\section{Results and Discussion}

In this section, we discuss the results of our study, examining the apparent resurgence of this newly identified Anonymous network relative to the renewed 2020 BLM protests. 

\subsection{Changes in the Anonymous Network: 2019 to 2020 (RQ1)}

In~\cite{Jones2020}, the authors found that Anonymous appeared to have fragmented as of December, 2019. They noted that the group had suffered a significant loss in accounts joining the network after 2013 (the year in which key Anonymous members were arrested) and that most accounts in the network were no longer active. In answer to RQ1, we examined the 2020 Anonymous network for evidence of an increase in network activity since 2019.

\begin{table*}[!ht]
\footnotesize
\centering
\begin{tabular}{p{0.25\linewidth} >{\raggedright\arraybackslash}p{0.65\linewidth}} 
 \toprule
 \textbf{Topics} & \textbf{Keywords} \\ [0.5ex] 
 \midrule
 Anonymous and BLM & anonymous, blacklivesmatter, legion, government, expect, follow, police, think, leak, forget \\
 \midrule
BLM Protests & black, life, matter, blacklivesmatter, street, georgefloyd, impunity, people, protest, icantbreathe \\
 \midrule
George Floyd Protests and BLM & icantbreathe, georgefloyd, blacklivesmatter, regime, trump, force, protestors, kpop, state, opfancam \\
  \midrule
George Floyd Protests & police, george, floyd, officer, murder, protest, justice, people, state, death \\
  \midrule
Police, Protest, and BLM & police, protester, cop, protest, officer, people, man, shot, blacklivesmatter, video \\
 \midrule
Police and Social Justice & woman, year, police, child, black, abuse, girl, man, killed, white \\
  \midrule
 Racial Tension & people, black, fuck, shit, fucking, white, racist, thing, guy, woman \\
   \midrule
Politics and Race & people, government, country, time, black, power, racism, change, life, police \\
  \midrule
US Presidential Election & trump, president, vote, biden, election, state, american, joe, republican, donald \\
 \midrule
 US Politics and China & trump, hong, kong, president, military, online, state, white, law, blacklivesmatter \\
  \midrule
International Politics & people, china, israel, state, palestinian, war, turkey, israeli, country, american \\
  \midrule
Social Issues & money, people, pay, school, work, worker, job, government, business, tax \\
  \midrule
COVID-19 & covid, people, case, death, covid19, mask, coronavirus, pandemic, virus, died \\
  \midrule
Computer Security & data, security, app, link, phone, government, user, tool, anonymous, file \\
 \midrule
 Julian Assange & assange, julian, journalist, anonymous, freeassange, people, crime, julianassange, prison, free \\
  \midrule
Positive Messaging & time, love, day, good, year, life, people, today, thing, friend \\
  \midrule
Anonymous and Fake Accounts & medium, account, fake, group, anonymous, campaign, real, people, bot, spread \\
  \midrule
Child Trafficking and Jeffrey Epstein & opdeatheaters, child, epstein, trump, trafficking, rape, crime, network, jeffrey, sex \\
  \midrule
Social Media & follow, facebook, retweet, day, instagram, video, read, post, people, time \\
 \bottomrule
\end{tabular}
\caption{Topics identified in tweets from the Anonymous Twitter network between 1st May and 31st August, 2020.}
\label{table:AnonTopics}
\end{table*}

\begin{figure}[!ht]
\centering
\includegraphics[width=0.6\linewidth]{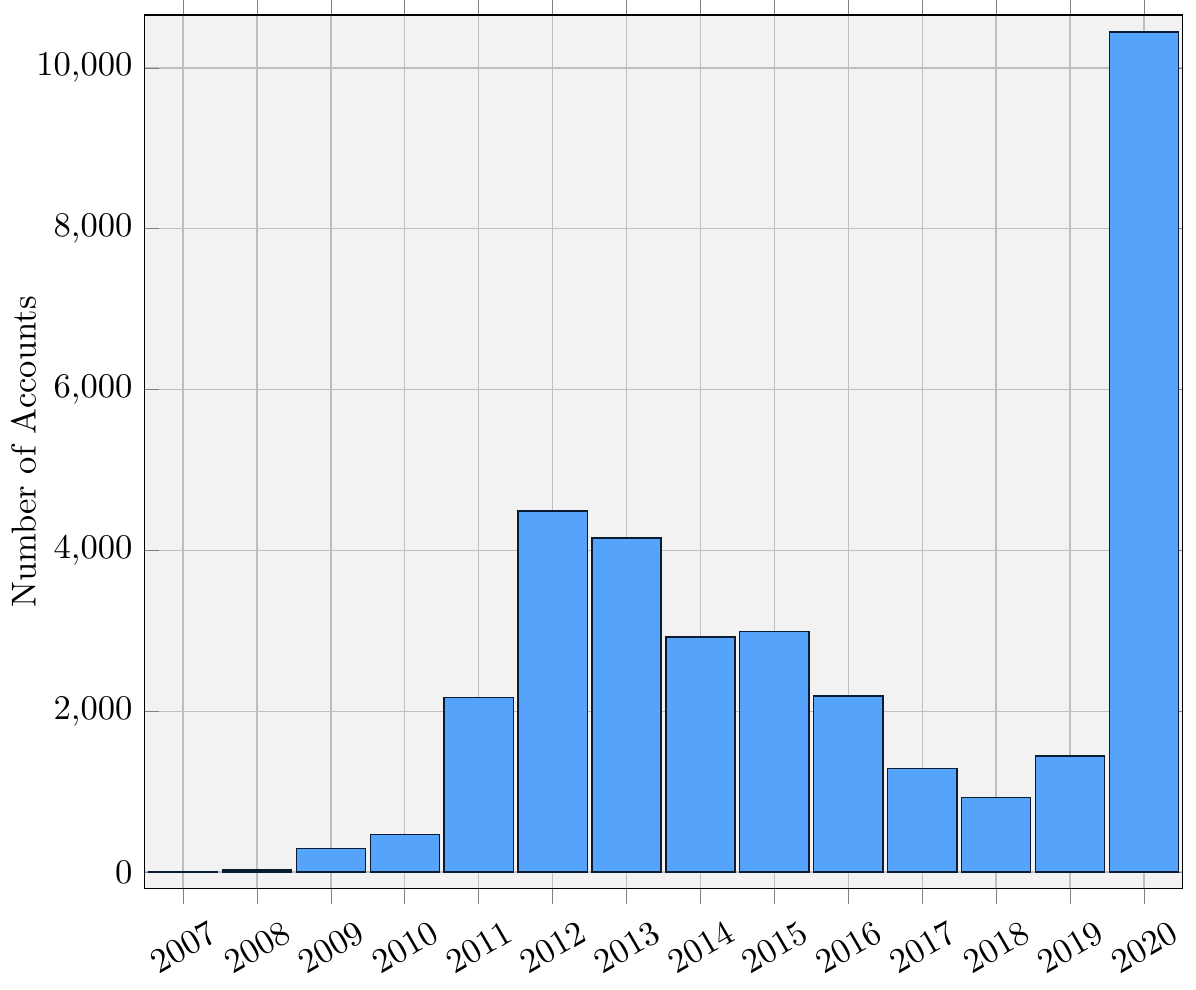}
\caption{The number of Anonymous accounts created in each year.}
\label{fig:createDate}
\end{figure}

In Fig.~\ref{fig:createDate}, we can see an immediate difference in the 2020 Anonymous network. In the year 2020, there is a substantial increase in the number of Anonymous accounts joining the network, a spike that is considerably greater than the group's largest gains in 2012. It therefore appears that a resurgence has occurred within the Anonymous network over the year 2020. 

To ensure that the difference in seed accounts (from~\cite{Jones2020}) was not harming our comparison, we utilized cosine similarity to examine the relationship between the number of accounts created each year between 2007 and 2019, for both the network found by~\citeauthor{Jones2020}, and our Anonymous network. This found a very high similarity of 0.99, offering strong indication that the difference in seeds did not significantly impact the sampling process.

Additionally, to ensure that the absence of deleted/suspended accounts was not artificially inflating the apparent difference between the number of accounts in 2020 compared to years prior, we identified the total number of removed accounts in our network. Although Twitter's API does not provide details regarding the reason for removal, nor any of the account's data prior to removal, it does notify API users that the account is no longer available. It was thus possible for us to approximate that the number of accounts removed was 1,184. Given the small number of accounts removed, and the significant difference in accounts created in 2020 when compared to other years, it is unlikely that the absence of these accounts has not materially impacted our findings.

\begin{figure}[!ht]
\centering
\includegraphics[width=0.6\linewidth]{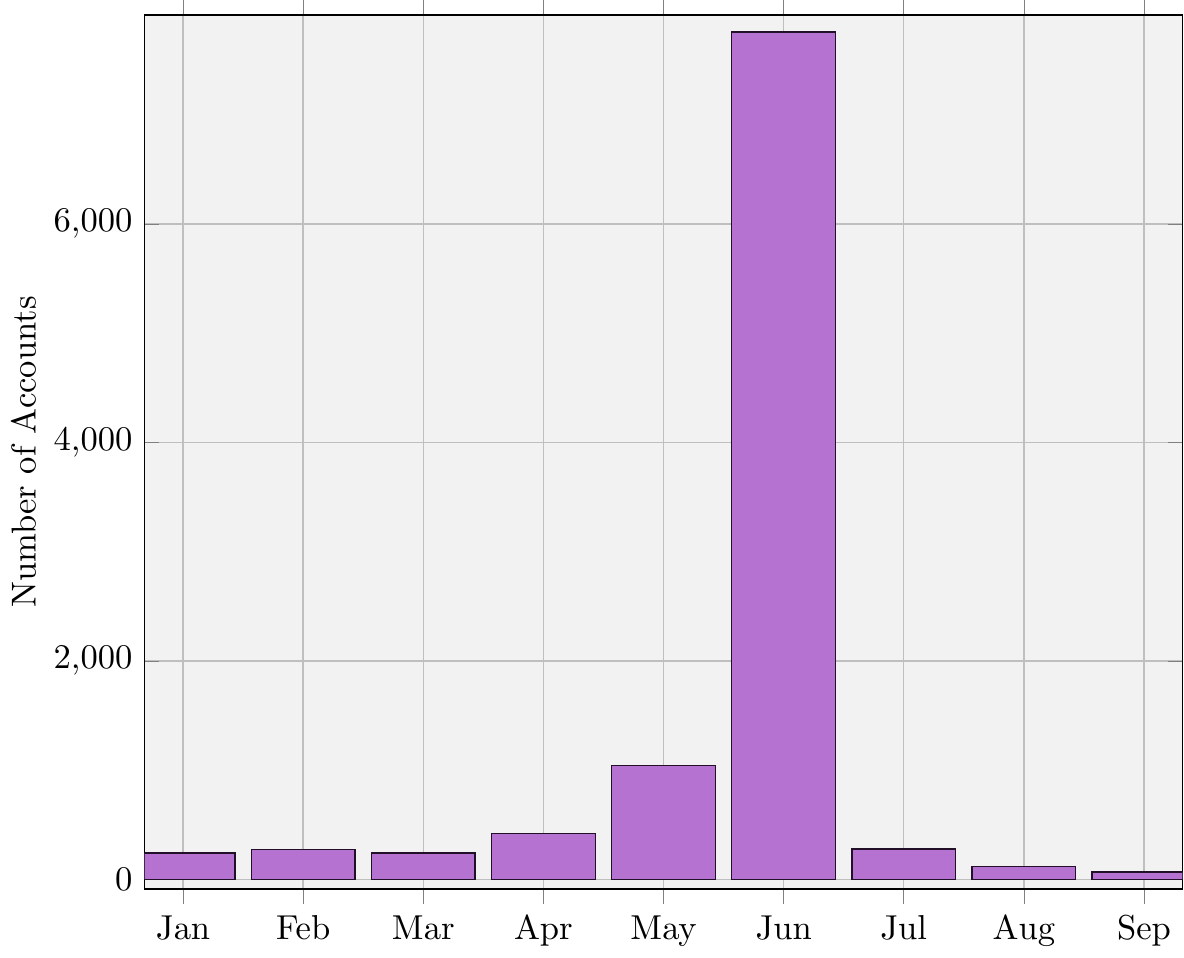}
\caption{The number of Anonymous accounts created in each month during the renewed BLM protests.}
\label{fig:createDates2020}
\end{figure}

To further examine this resurgence, we looked at the number of Anonymous accounts created per month in 2020 (Fig.~\ref{fig:createDates2020}). From this, we find a relatively stable set of accounts created from January to April, in line with the low number of new members that the Anonymous network had been receiving. In May, however, a notable uptick in new membership is recorded, followed by a large increase in June. An increase accounting for approximately 78\% of the total number of new members in the year so far. This dramatic increase, aligned with the time-period in which the BLM protests surged~\cite{NYT2020}, indicates that it may have been a driving factor in the growth of the Anonymous network.

From this, we can answer RQ1 in the affirmative. There appears to be significant evidence that a large-scale increase in accounts occurred not only in 2020, but specifically in the central months of the 2020 BLM protest.

\begin{figure*}[!htp]
\centering
\begin{subfigure}[b]{0.4\textwidth}
    \centering
    \includegraphics[width=0.95\textwidth]{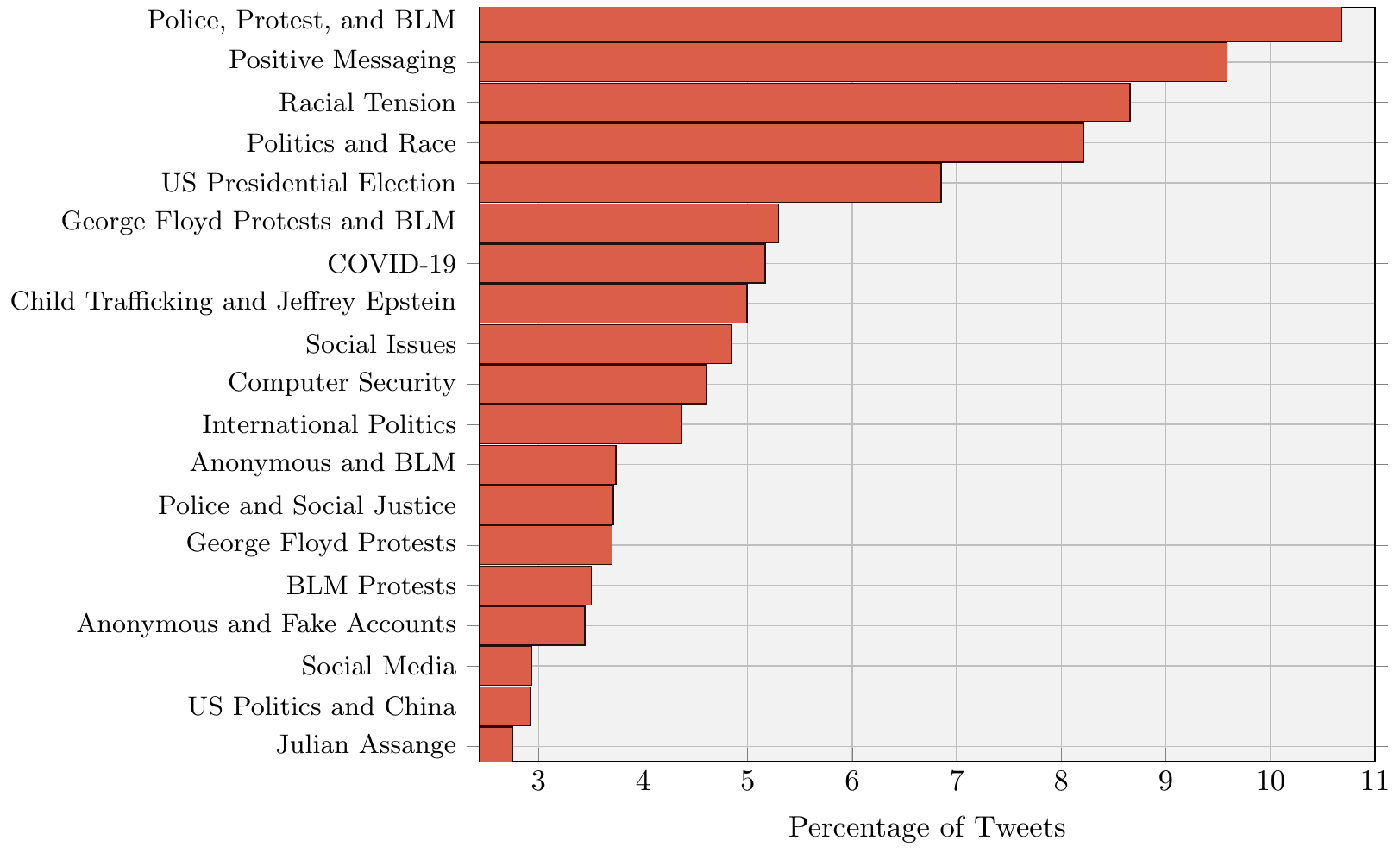}
    \caption{The percentage of Anonymous tweets containing each identified topic.}
    \label{fig:topicTweetPercs}
\end{subfigure}
\quad
\begin{subfigure}[b]{0.4\textwidth}
     \centering
     \includegraphics[width=0.9\textwidth]{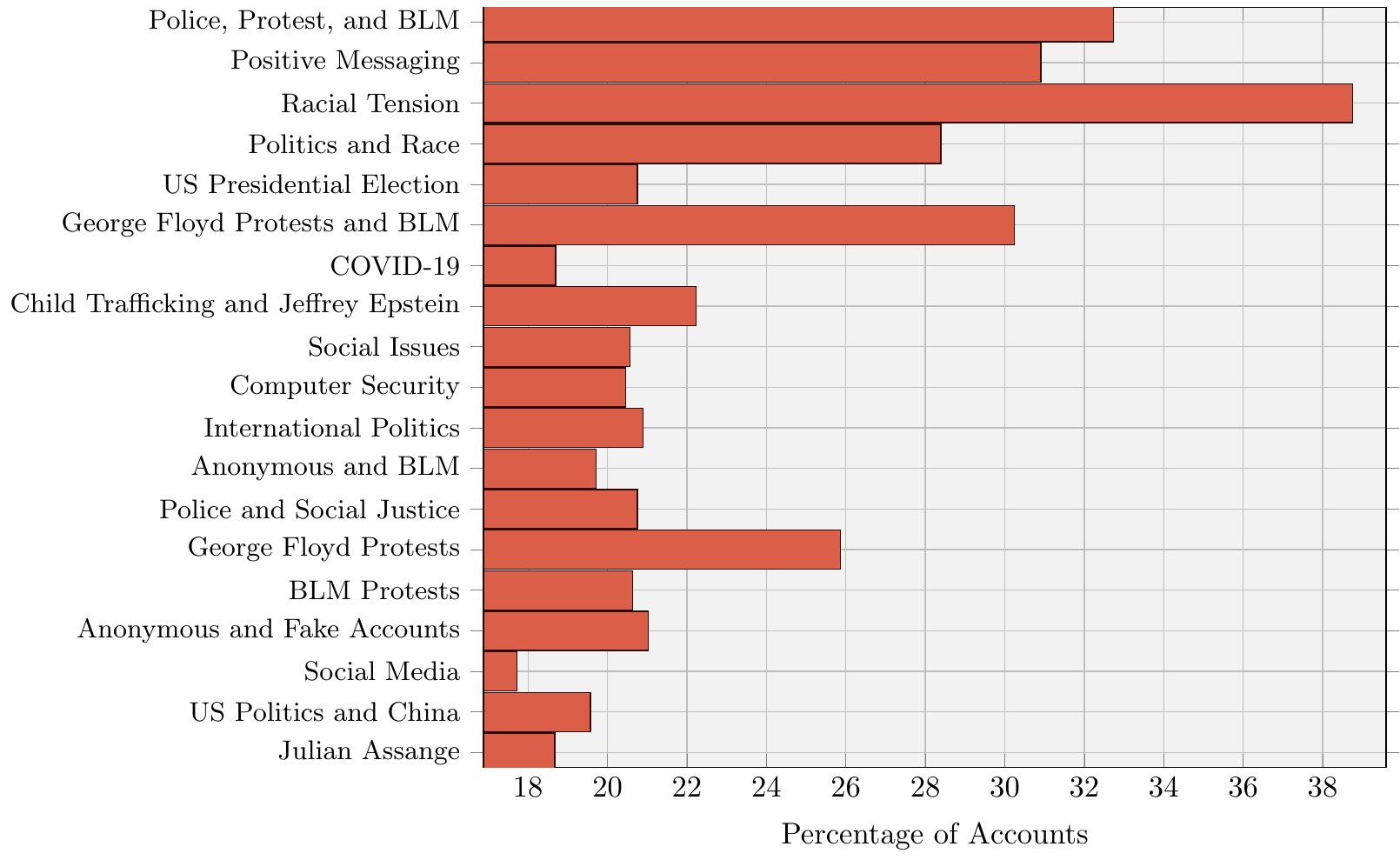}
     \caption{The percentage of Anonymous accounts tweeting about each topic.}
     \label{fig:topicAccountPercs}
\end{subfigure}
\caption{Topic frequencies in Anonymous tweets and Anonymous accounts with at least one tweet containing each topic, using a topic probability threshold of 0.8.}
\label{fig:topicPercs}
\end{figure*}

\subsection{Examining Topics of Discussion in Tweets from Anonymous Affiliates (RQ2, RQ3)}
\label{topic_analysis}

\begin{figure*}[!ht]
\centering
\begin{subfigure}[b]{0.4\textwidth}
    \centering
    \includegraphics[width=0.9\textwidth]{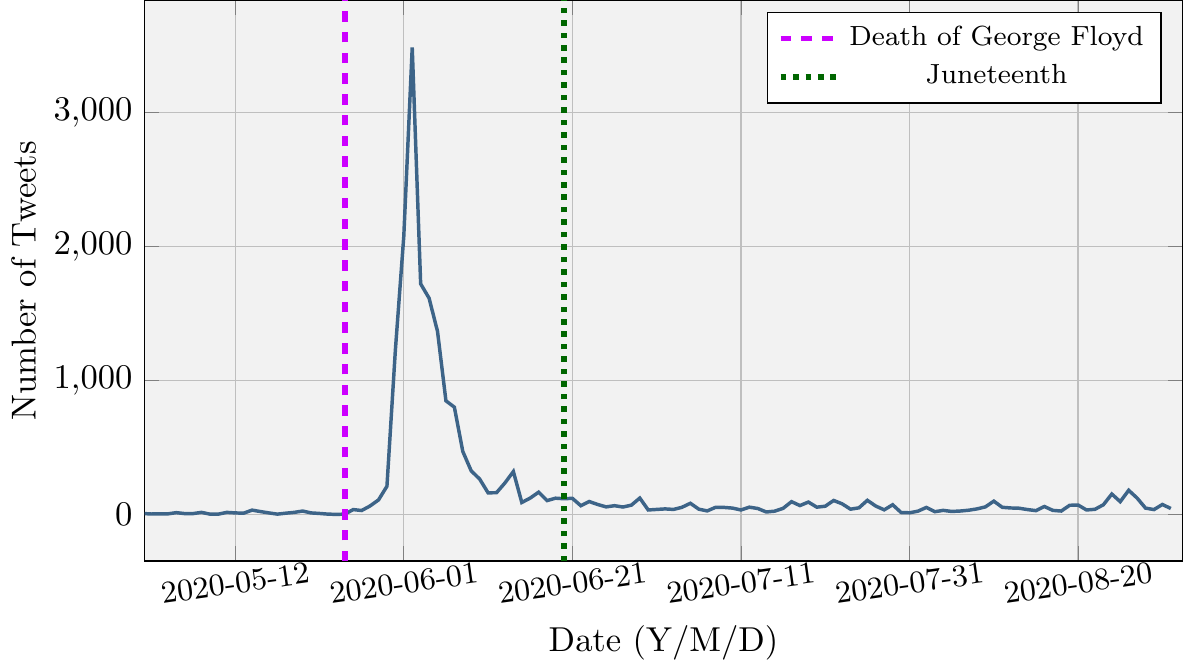}
    \caption{The number of BLM-focused tweets per day (1st May -- 31st August, 2020).}
\end{subfigure}
\quad
\begin{subfigure}[b]{0.4\textwidth}
     \centering
     \includegraphics[width=0.9\textwidth]{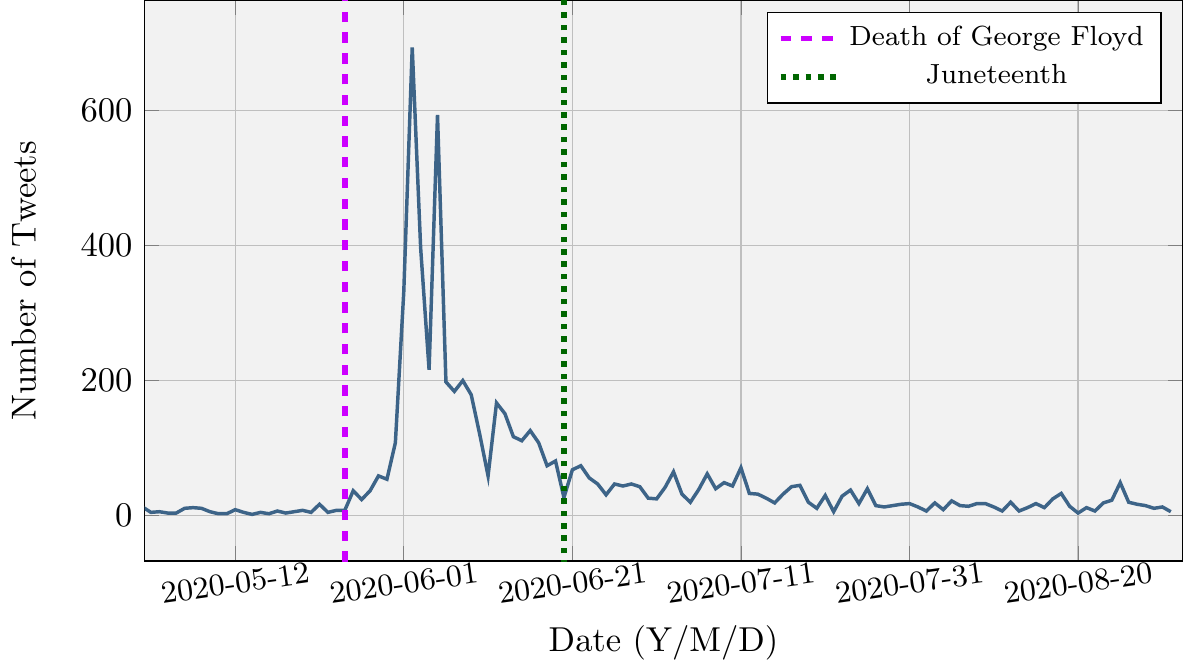}
     \caption{The number of police protest-focused tweets per day (1st May -- 31st August, 2020).}
\end{subfigure}
\caption{Graphs showing the number of tweets per day, between May and August, 2020, containing BLM and police protest related topics.}
\label{fig:TweetsPerDay}
\end{figure*}

Having identified a spike in activity in the Anonymous network around the 2020 BLM protests, we utilized topic modelling on tweets from the network created between the 1st May and 31st August, 2020. This, in turn, is used to answer RQs 2 and 3, examining the prominent topics during this period and the degree to which interest in these topics sustains itself after the period of sustained BLM protest in June, 2020.

The results of this analysis can be found in Table~\ref{table:AnonTopics}. Only 19 topics are shown, as the keywords for the remaining topic did not provide a clear indication of the subject they represented.

Most notable to our study, we find evidence of several topics pertaining to BLM, law enforcement, George Floyd-related protest, and racial justice. These findings are insightful, as they provide a broad indication that the Anonymous network, over the protest period, showed a clear interest in the BLM cause.

To further assess this, we examined the proportion of tweets in our dataset containing each topic (Fig.~\ref{fig:topicPercs}). From this, we found that of all tweets posted by Anonymous accounts created during the resurging spikes in May and June 2020, BLM-related topics were the most common (appearing in 26.95\% of tweets), with tweets considering more general topics of racial justice constituting a further 16.88\% of tweets. These topics appeared far more often than even the third most common topic, \textit{Positive Messaging} (9.59\%). Given that, based on the keywords learned for this topic, Positive Messaging presents itself as a broader, more general topic, this finding particularly emphasizes the central role that BLM protest topics played in tweets from the group during this period. This finding thus provides a direct link between the BLM protests and Anonymous' resurgence.

Alongside this, we also examined the percentage of accounts tweeting about each topic. The results for this can be found in Fig.~\ref{fig:topicAccountPercs}, where an account with at least one tweet about a given topic was considered to have tweeted about that topic. This revealed that BLM and race-related topics in particular are tweeted about by a sizeable proportion of the total number of accounts, with the \textit{Police, Protest, and BLM} topic being tweeted by 32.74\% of accounts, and the \textit{Racial Tension} topic, the topic with the highest proportion of unique accounts tweeting, it being tweeted by 38.76\% of Anonymous accounts. Moreover, for BLM/George Floyd-related topics, we observed that 63.7\% of accounts have tweeted at least one tweet regarding these topics. This further demonstrates that not only are BLM-related topics the most prominent in the network during this time, this prominence is not solely the result of a small number of highly active accounts. Instead, these topics appear to have been tweeted about by the majority of Anonymous accounts active during this period. Given the decentralized nature of the group and their claims of having no set interest or goals, this finding provides further confirmation that, beyond influencer accounts~\cite{Jones2020}, the Anonymous network as a whole presents a good deal of alignment in interests.

We then further examined the tweets identified as containing either topics pertaining to BLM, or police and protest. Any tweets belonging to either the \textit{Anonymous and BLM} topic, the \textit{BLM Protests} topic, the \textit{George Floyd and BLM} topic, or the \textit{Police, Protest, and BLM} topic were identified as BLM tweets. Moreover, any tweets belonging to either the \textit{Police and Social Justice} or the \textit{George Floyd Protests} topics were identified as police protest tweets.

From this, we examined the frequency of tweets per day for tweets belonging to each of these sets of topics. The results of which can be found in Fig.~\ref{fig:TweetsPerDay}. Through this analysis, we found evidence of a clear spike in activity for both topics following the death of George Floyd. Given that the foremost topics of interest in the Anonymous network are related to BLM and police protest more generally, these spikes provide good indication that interest in these topics were likely in response this this incident.

What we also see is that by the Juneteenth holiday, one of the last days of significant protest~\cite{NYT2020}, interest in the topic appears to have waned considerably, with very few tweets regarding either topic being posted. This is particularly surprising, given the sampling approach used. As our API usage required the sampling of the most recent tweets, one may expect that highly active accounts who had tweeted more than 3,200 times after the period of sustained protest could have disrupted these findings. However, even with this limitation in the sampling approach, this does not appear to be the case.

Thus, it seems that although the resurgence in Anonymous accounts is likely the result of the group's declared alliance with the BLM movement, this interest was short-lived, with the group very quickly losing interest after the period of sustained BLM protest. From this, we can confirm RQ2 -- there appears to be good evidence that BLM and wider topics of policing and racial injustice are both present in Anonymous tweets during this period, and are also popularly discussed. Additionally, we answer RQ3: it seems that although BLM-related topics are popularly discussed, there is little sustained interest in them after the BLM protests began to wane.

\subsection{Analyzing Anonymous' Response to the 2020 BLM Protests (RQ4)}

\begin{figure*}[!htp]
\centering
\begin{subfigure}[b]{0.45\textwidth}
    \centering
    \includegraphics[width=0.95\textwidth]{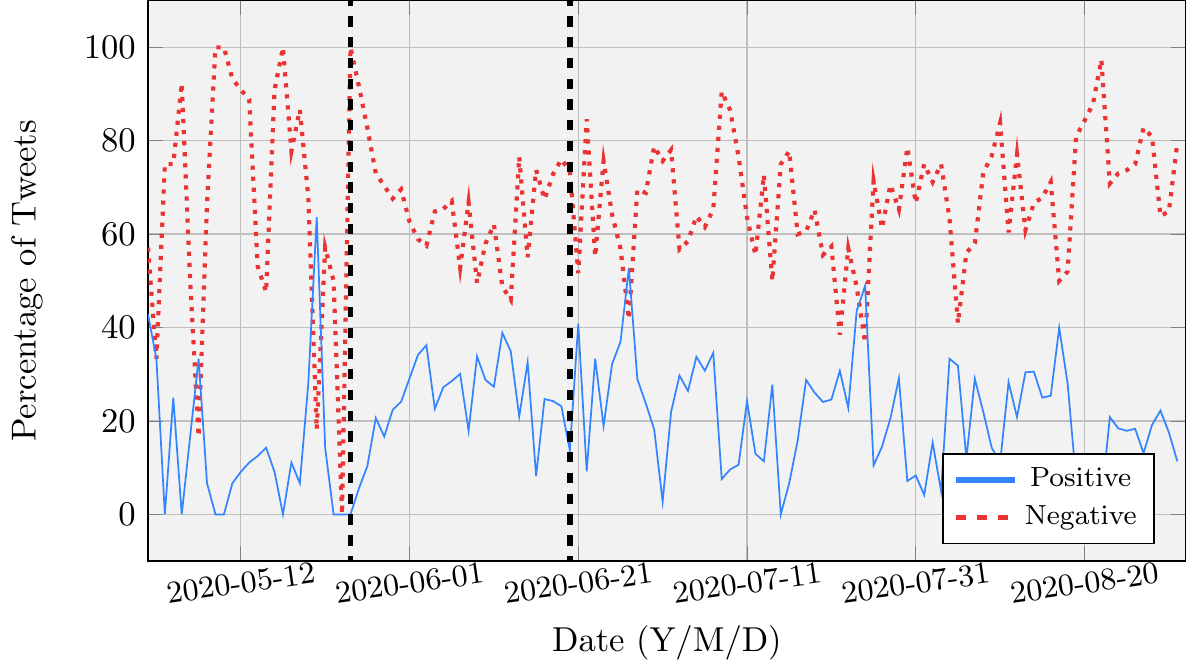}
    \caption{The proportion of positive and negative BLM tweets (1st May -- 31st August, 2020).}
    \label{fig:avgSentimentsBLM}
\end{subfigure}
\quad
\begin{subfigure}[b]{0.45\textwidth}
     \centering
     \includegraphics[width=0.95\textwidth]{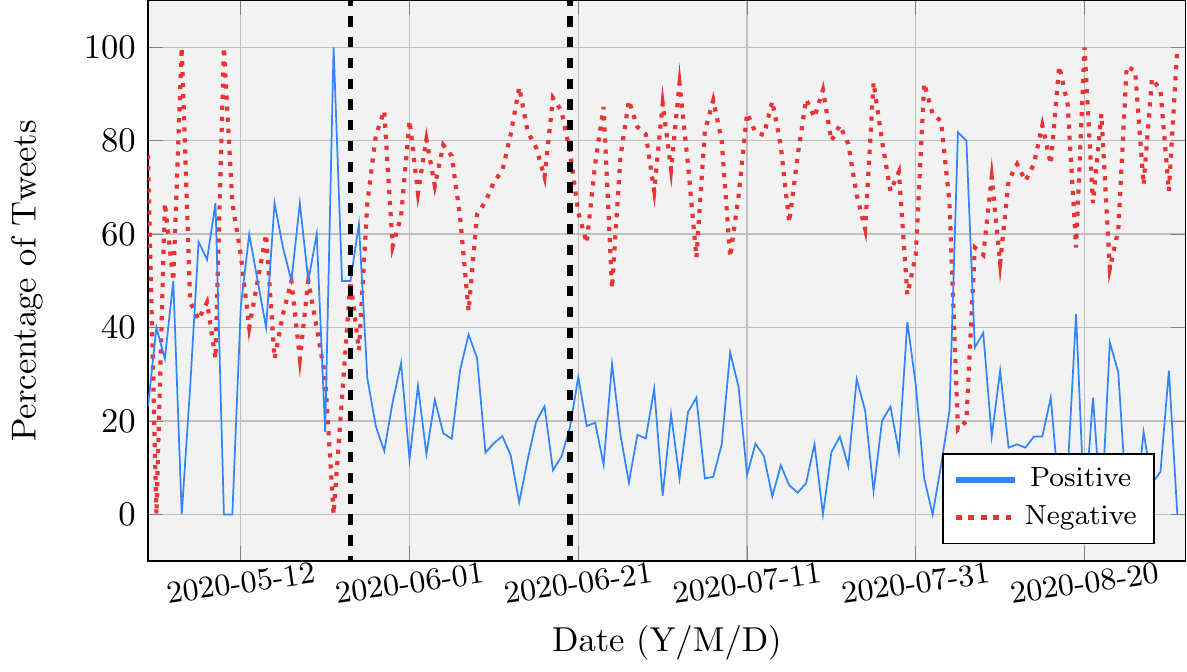}
     \caption{The proportion of positive and negative police protest tweets (1st May -- 31st August, 2020).}
     \label{fig:avgSentimentsPolice}
\end{subfigure}
\caption{The proportion of positive and negative Anonymous tweets containing BLM and police protest topics per day, during the 202 BLM protests. The first horizontal line denotes the death of George Floyd (May 25, 2020), and the second the Juneteenth holiday (June 19, 2020; the last day of notable protest in the US).}
\label{fig:avgSentimentsFiltered}
\end{figure*}

Beyond examining the topics of interest amongst tweets during the BLM protests, we were also interested in examining the sentiment of these tweets. In turn, we examined tweet sentiment in subsections of our tweet corpus, isolating tweets containing topics pertinent to the 2020 protests in the periods before, during, and after the peak in BLM protests in June. This, in turn, seeks to answer RQ4 -- providing indications of the manner in which Anonymous Twitter users engaged with BLM-related topics, and the degree to which unity existed amongst the group in relation to BLM. An important consideration given the group's claims to no set interests~\cite{Olson2013}, and their previous attacks against BLM in 2016~\cite{Verge2016}.

In Fig.~\ref{fig:avgSentimentsFiltered}, we show the sentiment change over time for tweets containing topics pertaining to BLM (Fig.~\ref{fig:avgSentimentsBLM}) and police protest (Fig.~\ref{fig:avgSentimentsPolice}). These tweets were identified in the same manner as applied in Section~\ref{topic_analysis}. For each tweet sentiment, the value was assigned to one of three classes: positive, negative, neutral, based on their VADER score -- with a score $>=0.05$ constituting a positive tweet, a score $<0.05$ and $>-0.05$ a neutral tweet, and a score $<-0.05$ a negative tweet. These values were chosen based on the recommendations of the tool's authors~\cite{Gilbert2014}.

We then used the chi-squared test of independence to study the relationship between tweet sentiment and time period. Comparing positive, negative, and neutral tweet sentiment frequencies before George Floyd's death and after his death but before the Juneteenth holiday, significant dependencies were identified between sentiment and time period for both BLM and police protest tweets ($\alpha=0.01$, $p<.001$ for all tests).

In Fig.~\ref{fig:avgSentimentsBLM}, we see that throughout the time period studied, tweet sentiment remains largely negative in topics discussing BLM. After George Floyd's death, however, we do see a narrowing between positive and negative sentiments. In Fig.~\ref{fig:avgSentimentsPolice}, the picture prior to George Floyd's death is quite sporadic, with frequent fluctuations between the proportions of positive and negative tweets. After his death, the pattern of sentiment changes quite rapidly, matching a similar trend in sentiment in the BLM-related tweets, with a steady proportion of negative tweets containing even after the period of increased protest.

To try and gain further understanding of what these changes in tweet sentiment mean in the context of their tweets, we manually examined BLM and police protest tweets of both positive and negative sentiment during each time period. What we found is that, typically, negative tweets containing these topics are being used to offer criticisms of police action, whilst positive tweets are documenting successes of the BLM-protests and the actions against police misconduct more broadly.

\begin{figure*}[!ht]
\centering
\begin{subfigure}[b]{0.45\textwidth}
     \centering
     \includegraphics[width=0.7\textwidth]{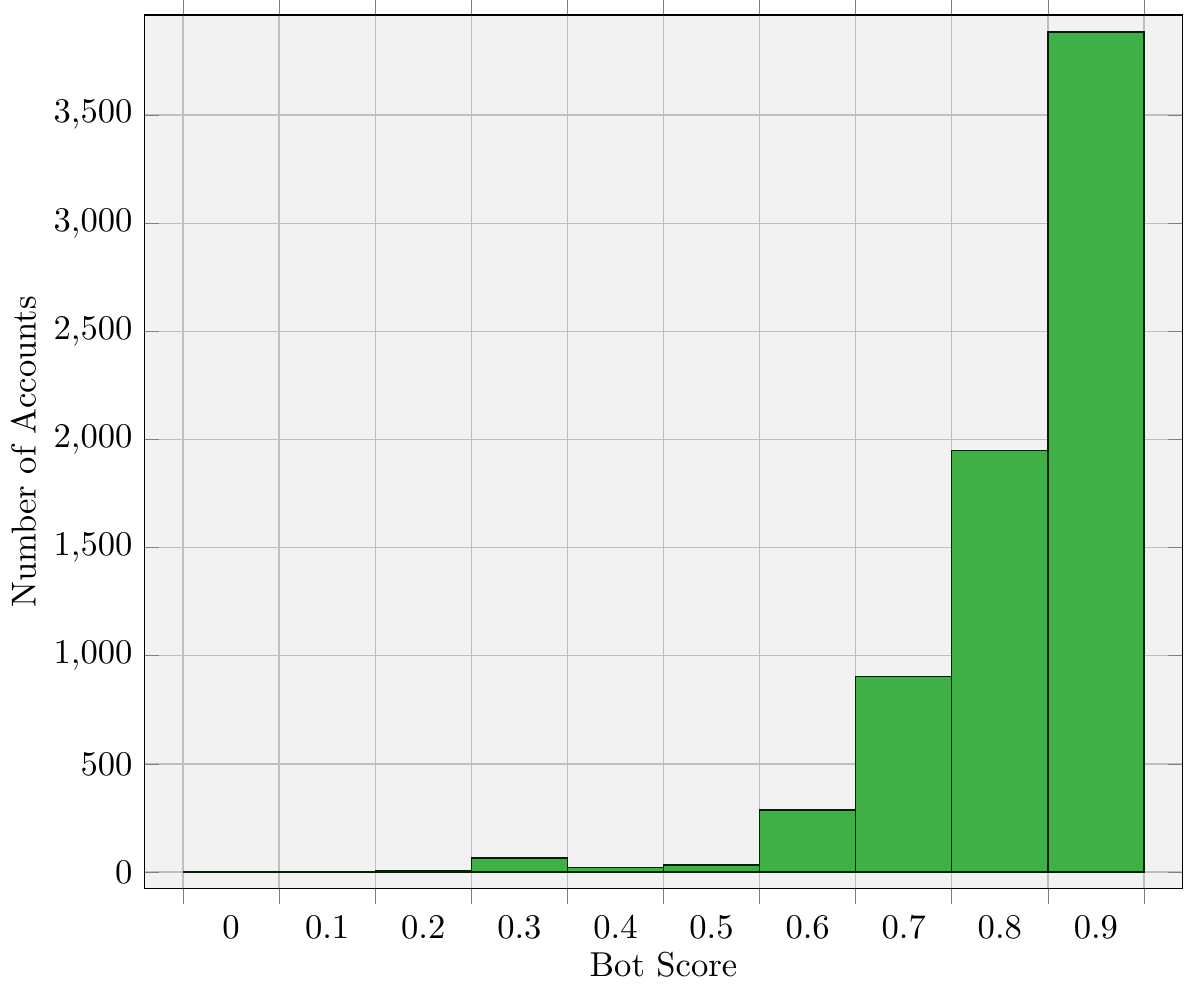}
     \caption{The distribution of bot scores in Anonymous Twitter network for accounts created in 2020.}
     \label{fig:botScores2020}
\end{subfigure}
\begin{subfigure}[b]{0.45\textwidth}
    \centering
    \includegraphics[width=0.7\textwidth]{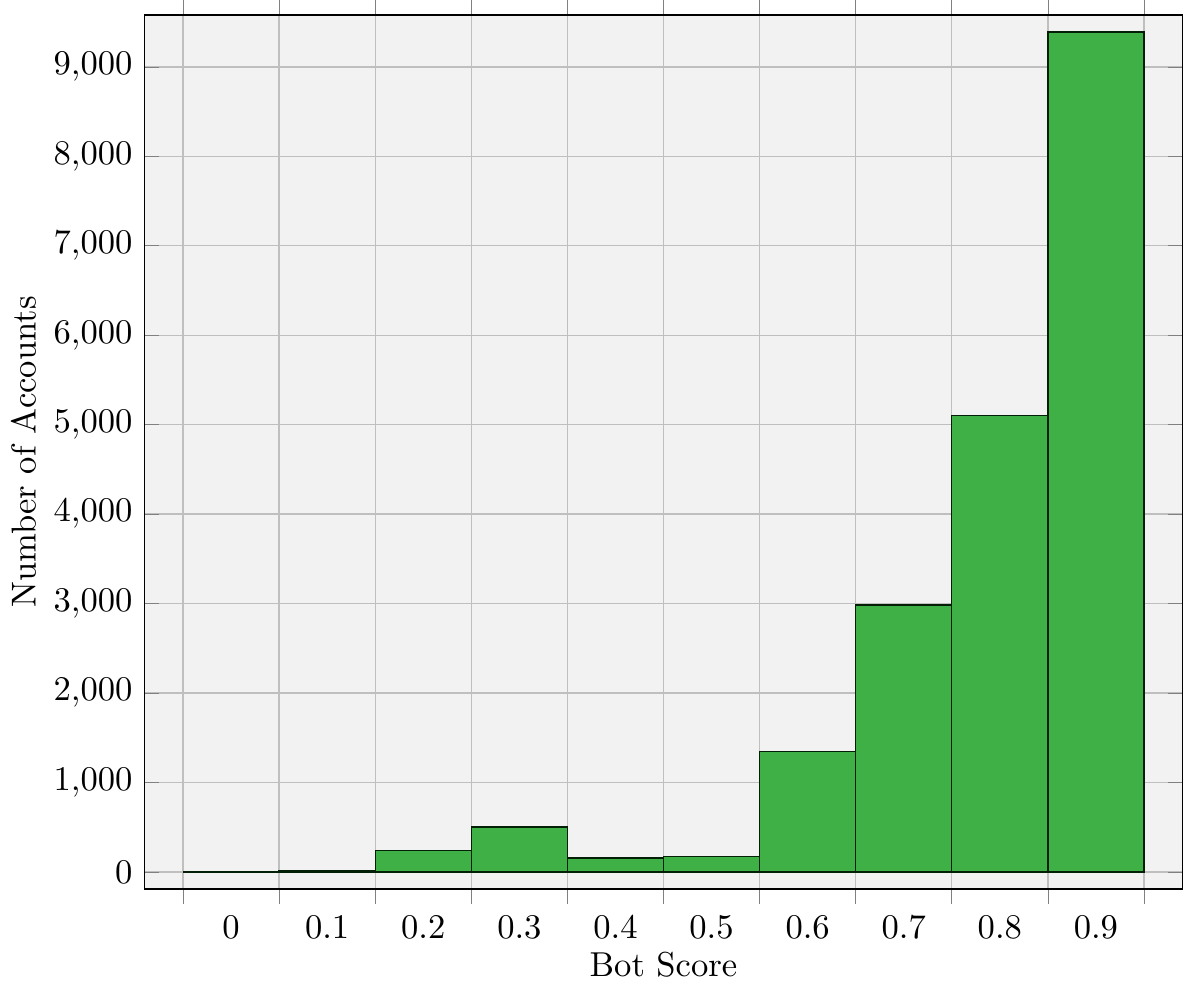}
    \caption{The distribution of bot scores in Anonymous Twitter network for accounts created prior to 2020.}
    \label{fig:botScores2019}
\end{subfigure}
\caption{The distributions of bot scores of Anonymous-affiliated Twitter accounts.}
\label{fig:botScores}
\end{figure*}

For instance, in BLM and police protest tweets prior to George Floyd's death, we see negative tweets such as ``\textit{This evening in Manhattan, an NYPD officer approaches a group of people filming an arrest, deploys his taser indiscriminately}'', ``\textit{\#BlackLivesMatter...tackled by police officers...\#PoliceBrutality}, and \textit{Black people get executed by police for just existing, while white people dressed like militia members carrying assault weapons are allowed to threaten State Legislators and staff...}''

In contrast, in positive tweets prior to George Floyd's murder we see examples such as ``\textit{But this will only grow our desire to fight their authoritarian and dictatorial regime. Anonymous is an idea; And they cannot arrest an idea!}'', ``\textit{...Federal judge orders ICE to release detainees from South Florida detention centers}'', and ``\textit{I'm calling on the Department of Justice to investigate. We need justice.}''

Typically, negative tweets are functioning to criticize the behaviours of police officers over a variety of different incidents, including the killings of Breonna Taylor and Ahmaud Arbery -- two other African Americans that were focal points for the resurgence of BLM actions~\cite{AlJazeera2020}. Whereas, whilst positive tweets have similar focuses, including the killing of Ahamud Arbery, they document activist actions and success stories, such as the sharing of footage of police brutality and appeals to governmental bodies for justice.

This also indicates that there does seem to have been some degree of support for BLM-related topics prior to George Floyd's death. Moreover, despite the fact that historically Anonymous has had an inconsistent relationship with BLM, attacking their websites in 2016~\cite{Verge2016}, the group, on Twitter at least, seems unified in their support for BLM-related causes prior to George Floyd's death, with no tweets criticizing BLM present in our dataset. With that being said, based on our findings in Fig.~\ref{fig:topicPercs}, although this level of support may have existed in the Anonymous network prior to George Floyd's death, these topics were discussed infrequently.

Moreover, this pattern of sentiment continues in tweets containing BLM and police protests topics during the period of significant protest after the death of George Floyd through to Juneteenth. Examples of negative tweets from this period include: ``\textit{The police shot an unarmed black man in the back...}'', ``\textit{...police...shot him...\#BlackLivesMatter}'', and ``\textit{A mass of troops or law enforcement is on the other side of the fence...How do you represent those who you fear? \#Anonymous \#DCProtest.}''

Whilst examples of positive tweets include: ``\textit{protect the protesters}'', ``\textit{Minneapolis council considers disbanding Police Department and replacing it...}'', and ``\textit{If you're protesting today please stay safe...\#blacklivesmatter}''.

Again, we typically see negative tweets being used as a means to criticize police action in a range of circumstances, and positive tweets as a means of expressing support for BLM. Given this knowledge, in Fig.~\ref{fig:avgSentimentsBLM} the increase in positive tweets after George Floyd's killing is likely a reflection of the increased BLM action, and thus the group's broadcasting of this action. Whilst the increased percentage of negative police protest tweets after George Floyd's killing indicates increased criticism of the police by Anonymous during this period.

Thus, in answer to RQ4, it seems that although a range of positive and negative sentiments exist across our studied timeline with respect to BLM topics, this typically represents the function of tweets as a means of expressing support for BLM, or criticizing police action. Anonymous, as a whole, therefore appears united on Twitter in their support for BLM. Thus, although significant shifts in typical tweet sentiments occur across each time period, these indicate changes in the ratio of these two tweet functions, rather than shifts in the stance of Anonymous users. It seems that, despite the group's claims of having no set stances or ideologies~\cite{Olson2013}, they appear united on Twitter in their support of BLM. This, in turn, also provides the first evidence of unity not just in topics of interest within the group, but also in the group's stance on the most popular topics being discussed.

\subsection{Bot Presence in the Anonymous Network (RQ5)}

Finally, in answer to RQ5, we examined the potential presence of automated accounts within the Anonymous network to examine the degree to which bots may have contributed to the resurgence of the group. In turn, we recorded bot scores for 27,059 Anonymous accounts. All scores are on a scale between 0 and 1, with a higher score indicating an account acting in a more automated manner.

In Fig.~\ref{fig:botScores}, we show the distribution of bot scores for accounts created in 2020 (Fig.~\ref{fig:botScores2020}), and accounts created prior to 2020 (Fig.~\ref{fig:botScores2019}). This allows us to examine whether any presence in automated behaviors has occurred after the spike in Anonymous activity in 2020

Our findings show that the network displays a high degree of automated behavior, with the majority of accounts scoring between 0.9 and 1.0. Moreover, over 50\% of all accounts scored above 0.8, indicating that the Anonymous network shows significant signs of bot-like behavior in most of its accounts. This finding helps bring in to doubt the true number of Anonymous users on Twitter. Interestingly, there appears to be little difference between the bot score distribution in 2020 accounts, and in accounts created prior to 2020. Instead, it appears that the majority of accounts in this Anonymous network have always exhibited a large degree of automated behavior.

\begin{figure}[!ht]
\centering
\includegraphics[width=0.6\linewidth]{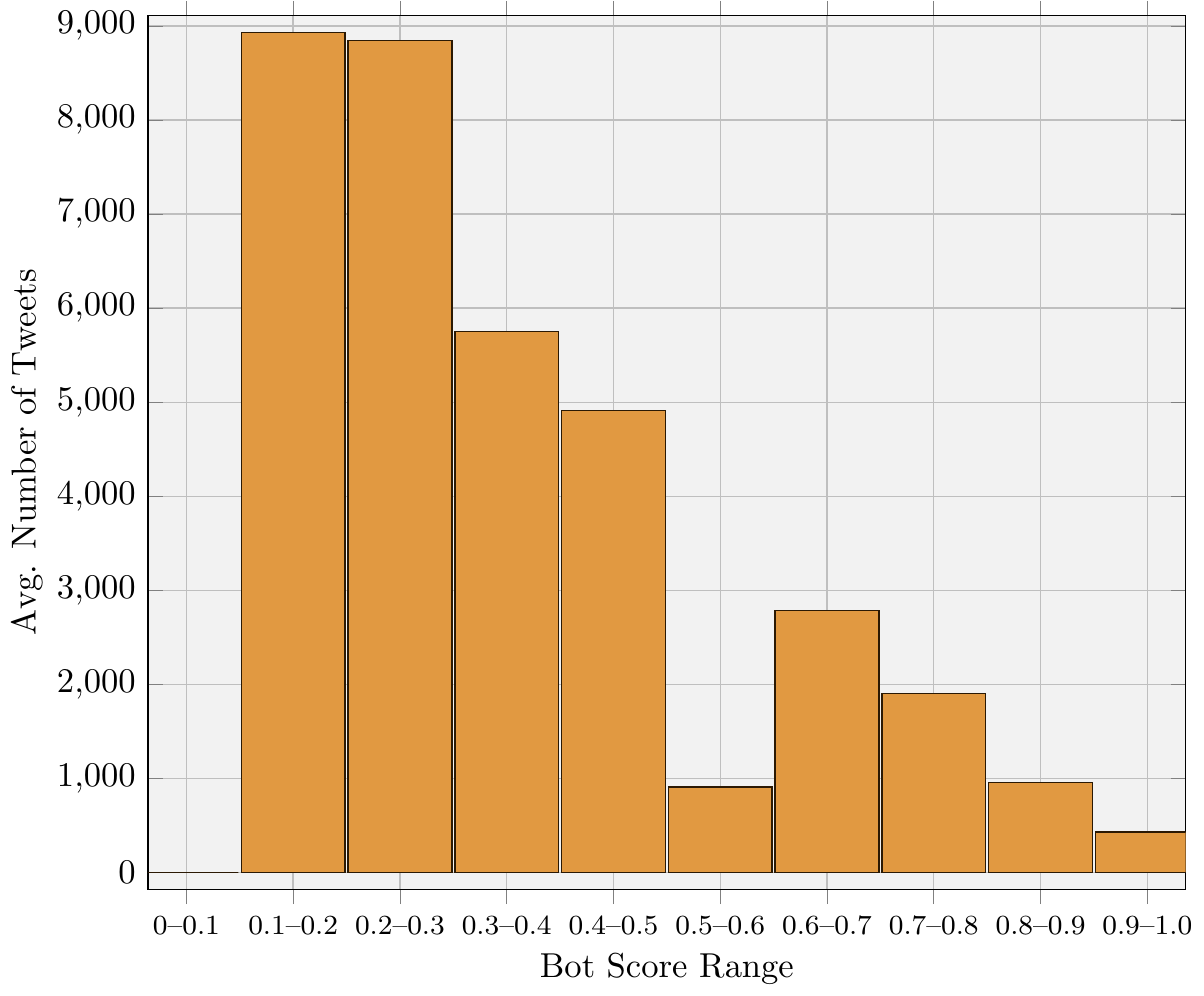}
\caption{The average number of total tweets for Anonymous accounts in different bot score ranges.}
\label{fig:botScoresTotalTweets}
\end{figure}

We then sought to examine the relationship between the number of tweets produced by Anonymous accounts and their bot scores. By doing this, we can begin to try and attain an understanding of whether there are any patterns in tweeting behaviors that distinguish accounts displaying automated patterns, and those seeming more human in nature. Fig.~\ref{fig:botScoresTotalTweets} shows the average number of tweets produced by accounts of different bot score ranges. In addition, we also examined the ratio of tweets to retweets in accounts with high and low bot scores, but this too did not identify any noticeable differences between more and less bot-like accounts.

From Fig.~\ref{fig:botScoresTotalTweets}, we can see that accounts scoring higher in terms of bot score seem, on average, to produce fewer tweets in total. This finding indicates that whilst accounts displaying automated behaviors appear to be present in large quantities within the Anonymous network, it seems that it is the accounts that act in the most human-like manner that are responsible for the majority of the tweets. Thus, whilst accounts exhibiting automated behaviors inflate the apparent size of the Anonymous network, they typically contribute little in the way of content.

Moving forward, we conducted more specific analysis into the relation between bot scores and BLM tweet topics. In turn, we examined the distribution of bot scores for accounts with at least tweet containing a BLM-related topic, and accounts with at least one police protest tweet. These distributions exhibited an even more considerable skew towards bot-like behavior, with all BLM and police protest tweeting accounts receiving a bot score of more than 0.8. This finding provides further indications that the apparent resurgence of the group in response to the BLM protests has been inflated by bot accounts.

\begin{figure}[!ht]
\centering
\includegraphics[width=0.6\linewidth]{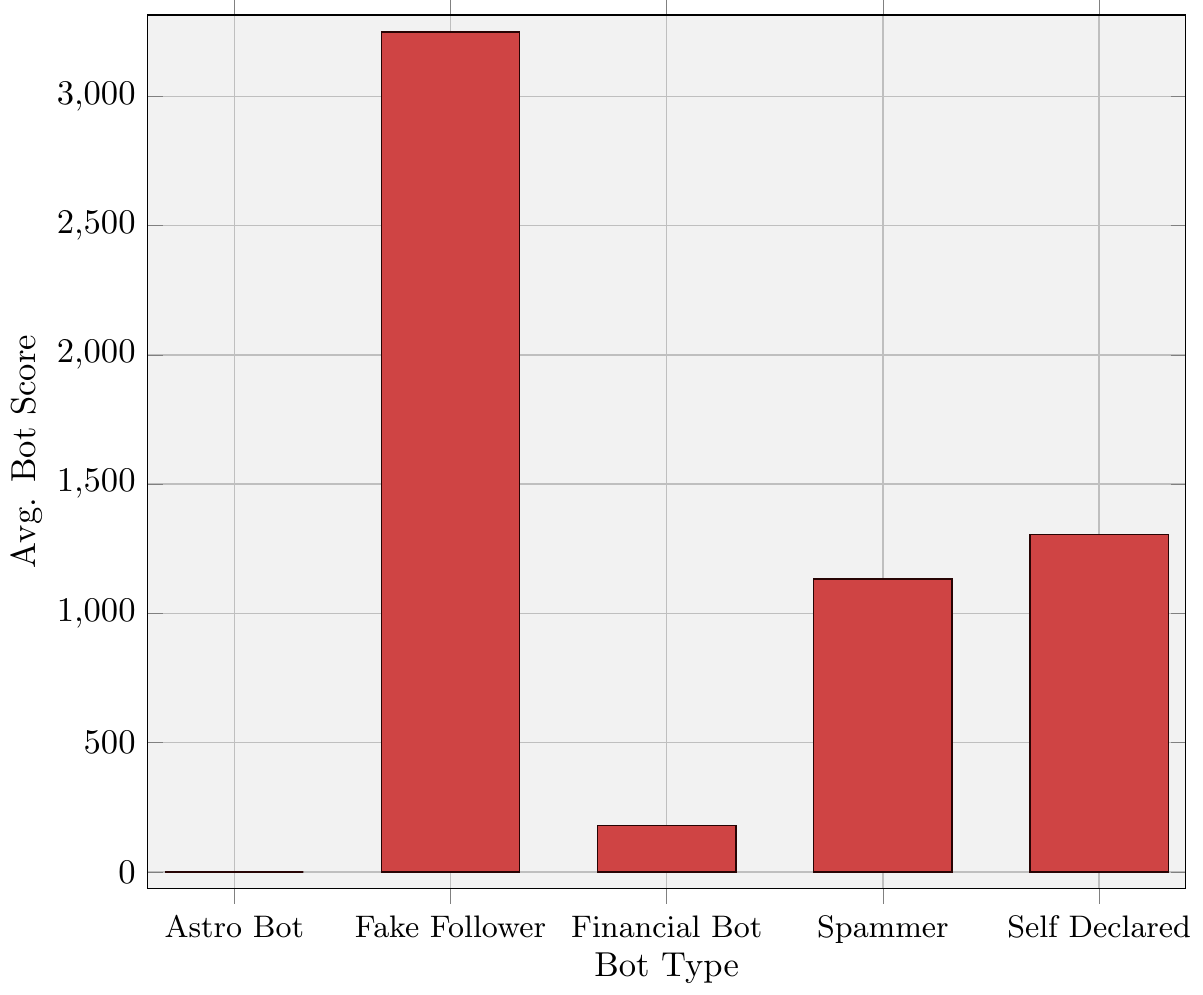}
\caption{The number of high-scoring Anonymous bot accounts created during the BLM protest period that exhibited behaviors of five specific bot types.}
\label{fig:botTypes}
\end{figure}

To further help confirm the role of these bot accounts in inflating the apparent size of the Anonymous network, we examined the bot type scores provided by Botometer. These provide an indication of the degree to which a given accounts behaves like a specific form of bot. The results can be found in Fig.~\ref{fig:botTypes}. From this, we can see that the `fake follower' bot type is the most commonly identified type present in Anonymous accounts created during the protest period. This provides additional evidence that the mass increase seen in the network during this period is likely the result of bot activity inflating the perceived size of the group's Twitter presence. 

Additionally, we conducted the same analysis for accounts created prior to 2020, finding similar indications that not only do accounts exhibiting bot-like behaviors constitute the majority of the network, but that the bots that they most commonly resemble are `fake follower'. It must be noted, however, that the majority of high-scoring bot accounts were left unclassified as `other' bots. Thus, further analysis is needed to complete our understanding of the role of bots in the network. 

However, these results, taken in conjunction with the other findings in regards to the presence of bots in the network, indicate, in answer to RQ5, that bot activity has played a significant role in inflating the apparent size of the Anonymous Twitter resurgence. We also find indications that the large network identified in \cite{Jones2020}, and Anonymous' sizeable presence on Twitter in general, is likely inflated by bot-like accounts. This finding is also in keeping with established behaviors of central affiliates within the group, as previous studies have noted that key affiliates have previously utilized bots as a means of inflating the group's apparent size to other affiliates~\cite{Olson2013}. This finding then, not only supports conjecture in the media that the rapid growth in Anonymous accounts during the 2020 BLM protests is likely suspicious in nature~\cite{Telegraph2020}, but also suggests that the Anonymous Twitter network has also been in large part construed of bot-like accounts prior to these events.

\subsection{Limitations}

There are some limitations to this study which warrant mentioning. Firstly, we make the assumption that each account is operated by a single user. In reality, it is possible that some accounts could be operated by a single user. An analysis of account behaviours for patterns of similarity might be possible to approximate the number of `true' users, though the lack of suitable ground-truth makes this a challenging problem.

Secondly, although we endeavoured to identify the number of removed accounts in the network, due to the limitations set by Twitter's API we cannot be certain that the number identified reflects the true number of removed accounts. However, given the similarity between our network and the one identified in~\cite{Jones2020}, coupled with the degree of growth in the network in 2020, it is unlikely that the loss of these accounts has severely impacted our findings.

Moreover, given our reliance on a generalizable bot detection method, further investigation is needed to validate our initial findings. Ideally, this would be done via the use of a bespoke classifier, trained specifically on Anonymous data. This is particularly needed in terms of the bot type detection.

Finally, due to the limitation of the most recent 3,200 tweets, our sampling is necessarily incomplete, and the results may be biased towards the accounts most active during the BLM protests. Validation of these results using a complete sampling would therefore be of value, as it could not be achieved in this work due to rate limit restrictions set by Twitter's API.

\section{Conclusions and Future Work}

In summary we find that, contrary to the findings of previous studies~\cite{Jones2020,Uitermark2017}, the group shows evidence of rapid growth in the time-period surrounding the 2020 BLM protests. Moreover, we find that the network as a whole frequently tweeted about BLM-related topics and that these tweets spiked considerably after George Floyd's death, supporting the notion that the Anonymous resurgence was, at least in part, a result of this. We also find indications, however, that this support was short-lived, with Anonymous showing little interest in BLM after the period of significant protest.

We also find evidence of automation across the majority of the accounts in the Anonymous network. This indicates that whilst the Anonymous network received a large amount of growth during the protests, much of this size may be the result of inflation through the use of bot accounts. Moreover, we note that bot accounts seem to constitute a large proportion of the Anonymous accounts that existed prior to 2020. This lends new insights into the group's presence on Twitter, indicating that the large presence of the group noted in past research~\cite{Jones2020} is likely not an accurate representation of the genuine number of Anonymous affiliates. 

Our results also indicate the potential power that bot activity has to mask the true extent of a group’s presence on social media. A finding which may have implications to the study of other groups with a significant online presence, such as QAnon, emphasizing a need for further research.

In future, the apparent role that automation plays in the Anonymous Twitter network could lead to a re-interpretation of the group's presence on social media. To strengthen our findings, we believe that further research into developing bespoke methods for identifying and analyzing bot activity in the Anonymous Twitter network would be valuable. Furthermore, analysis of the interaction between bot and human accounts would be of interest and would lend further context to the role that bot accounts play in the Anonymous network.


\fontsize{9.0pt}{10.0pt} 
\selectfont
\bibliography{FULL-JonesK} 

\begin{thebibliography}{22}
\providecommand{\natexlab}[1]{#1}
\providecommand{\url}[1]{\texttt{#1}}
\providecommand{\urlprefix}{URL }
\expandafter\ifx\csname urlstyle\endcsname\relax
  \providecommand{\doi}[1]{doi:\discretionary{}{}{}#1}\else
  \providecommand{\doi}{doi:\discretionary{}{}{}\begingroup
  \urlstyle{rm}\Url}\fi

\bibitem[{{Al Jazeera Media Network}(2020)}]{AlJazeera2020}
{Al Jazeera Media Network}. 2020.
\newblock A timeline of the {George Floyd} and anti-police brutality protests.
\newblock Al Jazeera.
\newblock
  \urlprefix\url{https://www.aljazeera.com/news/2020/06/11/a-timeline-of-the-george-floyd-and-anti-police-brutality-protests/}.
\newblock Accessed: 05-15-2021.

\bibitem[{Beraldo(2017)}]{Beraldo2017}
Beraldo, D. 2017.
\newblock \emph{Contentious Branding: Reassembling Social Movements Through
  Digital Mediators}.
\newblock Ph.D. thesis, University of Amsterdam.
\newblock
  \urlprefix\url{https://dare.uva.nl/search?identifier=a293284c-257a-43d2-a26d-e5cb3eaa4e8d}.

\bibitem[{Brandom(2016)}]{Verge2016}
Brandom, R. 2016.
\newblock {Anonymous} groups attacked {Black Lives Matter} website for six
  months.
\newblock The Verge.
\newblock
  \urlprefix\url{https://www.theverge.com/2016/12/14/13951762/anonymous-black-lives-matter-ddos-attack-six-months-hacktivism}.
\newblock Accessed: 05-15-2021.

\bibitem[{Buchanan(2020)}]{NYT2020}
Buchanan, L. 2020.
\newblock {Black Lives Matter} May Be the Largest Movement in {U.S.} History.
\newblock The New York Times.
\newblock
  \urlprefix\url{https://www.nytimes.com/interactive/2020/07/03/us/george-floyd-protests-crowd-size.html}.
\newblock Accessed: 05-15-2021.

\bibitem[{Burns(2020)}]{AJC2020}
Burns, A.~S. 2020.
\newblock {Atlanta} police site goes down, hacker group claims responsibility.
\newblock The Atlanta Journal-Constitution.
\newblock
  \urlprefix\url{https://www.ajc.com/news/breaking-news/hacker-group-claims-responsibility-after-apd-website-goes-offline/pYHBqHERBqPYqzPuC5tTxL/}.
\newblock Accessed: 05-15-2021.

\bibitem[{Carney(2016)}]{Carney2016}
Carney, N. 2016.
\newblock All Lives Matter, But So Does Race: {Black Lives Matter} and the
  Evolving Role of Social Media.
\newblock \emph{Humanity \& Society} 40(2): 180--199.

\bibitem[{Ferrara(2020)}]{Ferrara2020}
Ferrara, E. 2020.
\newblock What types of {COVID}-19 conspiracies are populated by {Twitter}
  bots?
\newblock \emph{First Monday} 25(6).

\bibitem[{Griffin(2020)}]{Independant2020}
Griffin, A. 2020.
\newblock `{Anonymous}' Activists Return With Hugely Popular Messages of
  Support for {George Floyd} Protests.
\newblock Independent.
\newblock
  \urlprefix\url{https://www.independent.co.uk/life-style/gadgets-and-tech/news/anonymous-george-floyd-black-lives-matter-facebook-twitter-video-k-pop-a9542666.html}.
\newblock Accessed: 05-15-2021.

\bibitem[{Hutto and Gilbert(2014)}]{Gilbert2014}
Hutto, C.~J.; and Gilbert, E. 2014.
\newblock {VADER}: A Parsimonious Rule-Based Model for Sentiment Analysis of
  Social Media Text.
\newblock In \emph{Proceedings of the 8th International AAAI Conference on
  Weblogs and Social Media}, 216--225. AAAI.

\bibitem[{Jones, Nurse, and Li(2020)}]{Jones2020}
Jones, K.; Nurse, J.~R.~C.; and Li, S. 2020.
\newblock Behind the Mask: A Computational Study of {Anonymous}' Presence on
  {Twitter}.
\newblock In \emph{Proceedings of the 14th International AAAI Conference on Web
  and Social Media}, 327--338. AAAI.

\bibitem[{Kigerl(2018)}]{Kigerl2018}
Kigerl, A. 2018.
\newblock Profiling Cybercriminals: Topic Model Clustering of Carding Forum
  Member Comment Histories.
\newblock \emph{Social Science Computer Review} 36(5): 591--609.

\bibitem[{McGovern and Fortin(2020)}]{McGovern2020}
McGovern, V.; and Fortin, F. 2020.
\newblock The {Anonymous} Collective: Operations and Gender Differences.
\newblock \emph{Women and Criminal Justice} 30(2): 91--105.

\bibitem[{Murdock(2016)}]{IBT2016}
Murdock, J. 2016.
\newblock {Anonymous} declares `day of solidarity' with {Black Lives Matter} to
  protest police brutality.
\newblock International Business Times.
\newblock
  \urlprefix\url{https://www.ibtimes.co.uk/anonymous-declares-day-solidarity-black-lives-matter-protest-police-brutality-1569983}.
\newblock Accessed: 05-15-2021.

\bibitem[{Murphy(2020)}]{Telegraph2020}
Murphy, M. 2020.
\newblock Hacking group {Anonymous} has returned for {George Floyd} protests
  –- or has it?
\newblock The Telegraph.
\newblock
  \urlprefix\url{https://www.telegraph.co.uk/technology/2020/06/05/hacking-group-anonymous-has-returned-george-floyd-protests/}.
\newblock Accessed: 05-15-2021.

\bibitem[{Olson(2013)}]{Olson2013}
Olson, P. 2013.
\newblock \emph{We Are {Anonymous}}.
\newblock William Heinemann.
\newblock ISBN 9780434022083.

\bibitem[{Pedregosa et~al.(2011)Pedregosa, Varoquaux, Gramfort, Michel,
  Thirion, Grisel, Blondel, Prettenhofer, Weiss, Dubourg, Vanderplas, Passos,
  Cournapeau, Brucher, Perrot, and Duchesnay}]{Pedregosa2011}
Pedregosa, F.; Varoquaux, G.; Gramfort, A.; Michel, V.; Thirion, B.; Grisel,
  O.; Blondel, M.; Prettenhofer, P.; Weiss, R.; Dubourg, V.; Vanderplas, J.;
  Passos, A.; Cournapeau, D.; Brucher, M.; Perrot, M.; and Duchesnay, E. 2011.
\newblock {Scikit-learn}: Machine Learning in {Python}.
\newblock \emph{Journal of Machine Learning Research} 12: 2825--2830.

\bibitem[{Rauchfleisch and Kaiser(2020)}]{Rauchfleisch2020}
Rauchfleisch, A.; and Kaiser, J. 2020.
\newblock The false positive problem of automatic bot detection in social
  science research.
\newblock \emph{PLOS ONE} 15(10).

\bibitem[{{\v R}eh{\r u}{\v r}ek and Sojka(2010)}]{Gensim}
{\v R}eh{\r u}{\v r}ek, R.; and Sojka, P. 2010.
\newblock Software Framework for Topic Modelling with Large Corpora.
\newblock In \emph{Proceedings of the LREC 2010 Workshop on New Challenges for
  NLP Frameworks}, 45--50.

\bibitem[{Syed and Spruit(2017)}]{Syed2017}
Syed, S.; and Spruit, M. 2017.
\newblock Full-text or Abstract? Examining Topic Coherence Scores Using Latent
  {Dirichlet} Allocation.
\newblock In \emph{Proceedings of the 2017 IEEE International Conference on
  Data science and Advanced Analytics}, 165--174. IEEE.

\bibitem[{{Twitter, Inc.}(2020)}]{TwitterAPI}
{Twitter, Inc.} 2020.
\newblock {Twitter} {API}.
\newblock Web page.
\newblock \urlprefix\url{https://developer.twitter.com/en/docs/twitter-api}.
\newblock Accessed: 05-15-2021.

\bibitem[{{Twitter, Inc.}(2021)}]{TwitterDev}
{Twitter, Inc.} 2021.
\newblock {Twitter} Developer Policy.
\newblock Web page.
\newblock
  \urlprefix\url{https://developer.twitter.com/en/developer-terms/policy#c-respect-users-control-and-privacy}.
\newblock Accessed: 05-15-2021.

\bibitem[{Uitermark(2017)}]{Uitermark2017}
Uitermark, J. 2017.
\newblock Complex Contention: Analyzing Power Dynamics Within {Anonymous}.
\newblock \emph{Social Movement Studies} 16(4): 403--417.

\end{thebibliography}

\end{document}